\documentclass[aps, prb, twocolumn, amsmath, amssymb, reprint, groupedaddress]{revtex4-1}
\usepackage{graphicx, hyperref, color}

\begin{document}

\title{Spin dynamical phase and anti-resonance in a strongly coupled magnon-photon system}

\author{Michael Harder}
\email{michael.harder@umanitoba.ca}

\author{Paul Hyde}

\author{Lihui Bai}
\email{bai@physics.umanitoba.ca}

\author{Christophe Match}

\author{Can-Ming Hu}

\affiliation{Department of Physics and Astronomy, University
of Manitoba, Winnipeg, Canada R3T 2N2}
 
\date{\today}

\begin{abstract}
We experimentally studied a strongly coupled magnon-photon system via microwave transmission measurements.  An anti-resonance, i.e. the suppression of the microwave transmission, is observed, indicating a relative phase change between the magnon response and the driving microwave field.  We show that this anti-resonance feature can be used to interpret the phase evolution of the coupled magnon-microwave system and apply this technique to reveal the phase evolution of magnon dark modes. Our work provides a standard procedure for the phase analysis of strongly coupled systems, enabling the phase characterization of each subsystem, and can be generally applied to other strongly coupled systems.  
\end{abstract}

\maketitle

\section{Introduction}

Very recently magnon-photon systems comprised of a ferrimagnetic material strongly coupled to a microwave cavity mode have been intensively studied via  microwave transmission measurements \cite{Huebl2013a, Tabuchi2014a, Zhang2014,Goryachev2014,Bhoi2014, Bai2015}.  Analysis of the microwave transmission \textit{amplitude} has revealed tell-tale signatures of strong coupling, such as normal mode anti-crossing and damping evolution, with such features accurately described by several theoretical approaches, including a linear quantum description \cite{Zhang2014}, a classical electrodynamic theory \cite{Bai2015, Cao2014}, a microwave transfer matrix approach \cite{Yao2015}, as well as a coupled harmonic oscillator model \cite{Harder2016}.  Due to their experimental simplicity and potential for both hybrid quantum information and spintronic applications, these coupled magnon-photon systems have immediately attracted much attention \cite{Lambert2015,Haigh2015a,Yao2015a,Abdurakhimov2015, Bai2016}.  Examples of early applications include the design of magnon dark mode information storage architectures \cite{Zhang2015g} and the use of strong coupling as a bridge between ferromagnetic resonance (FMR) and qubit systems \cite{Tabuchi2015b}.  In both of these examples the nature of the correlated magnon and photon phases plays a key role in the desired coherent information transfer.  Therefore a better understanding of the magnetization and cavity mode \textit{phases}, as well as a robust detection method to determine such phase information, is necessary for the future development of coherent information processing based on strongly coupled spin-photon systems. 

An intriguing possibility is to characterize phase information via anti-resonance phenomena.  It is well known that changes in the relative phase between the two components of a coupled system can lead to destructive interference between an external driving force on one component and the feedback force generated by the other component.  This so called anti-resonance phenomena leads to a suppressed resonance amplitude, rather than the typical enhanced amplitude seen at resonance frequencies.  Such anti-resonance features have previously been observed in a wide range of physical systems, from light scattering in meta-materials \cite{Rybin2015} to coupled atom-photon systems \cite{Sames2014}, and have been used to characterize strongly coupled quantum circuits \cite{Sames2014}, where the system components can not otherwise be characterized individually.  Yet despite the useful role of anti-resonance behaviour in strongly coupled systems, such features have not been investigated in the context of strongly coupled spin-photon systems.

In this work we studied a strongly coupled spin-photon system, formed by the ferrimagnetic insulator yttrium-iron-garnet (YIG) and a microwave cavity, observing an anti-resonance in the microwave transmission.  As the anti-resonance solely depends on the FMR we can characterized the uncoupled FMR properties, such as the resonance frequency, damping and $\omega-H$ dispersion, despite the fact that the FMR is strongly coupled to the cavity mode.  At anti-resonance a phase shift is observed in the microwave transmission spectra.  Using a classical electrodynamic description of the spin-photon system, the phase information of each subsystem can be determined, with the observed anti-resonance phase shift naturally appearing.  Our work therefore introduces a standard procedure for the analysis of individual components in a strongly coupled system, provides a method to extract the phase information of both the spin and cavity subsystems, and offers a natural interpretation of the observed phase shifts based on the concept of anti-resonance. 

\section{Microwave Transmission of the Coupled Magnon-Photon System}

\begin{figure}[tb]
 \includegraphics[width = 7.70 cm]{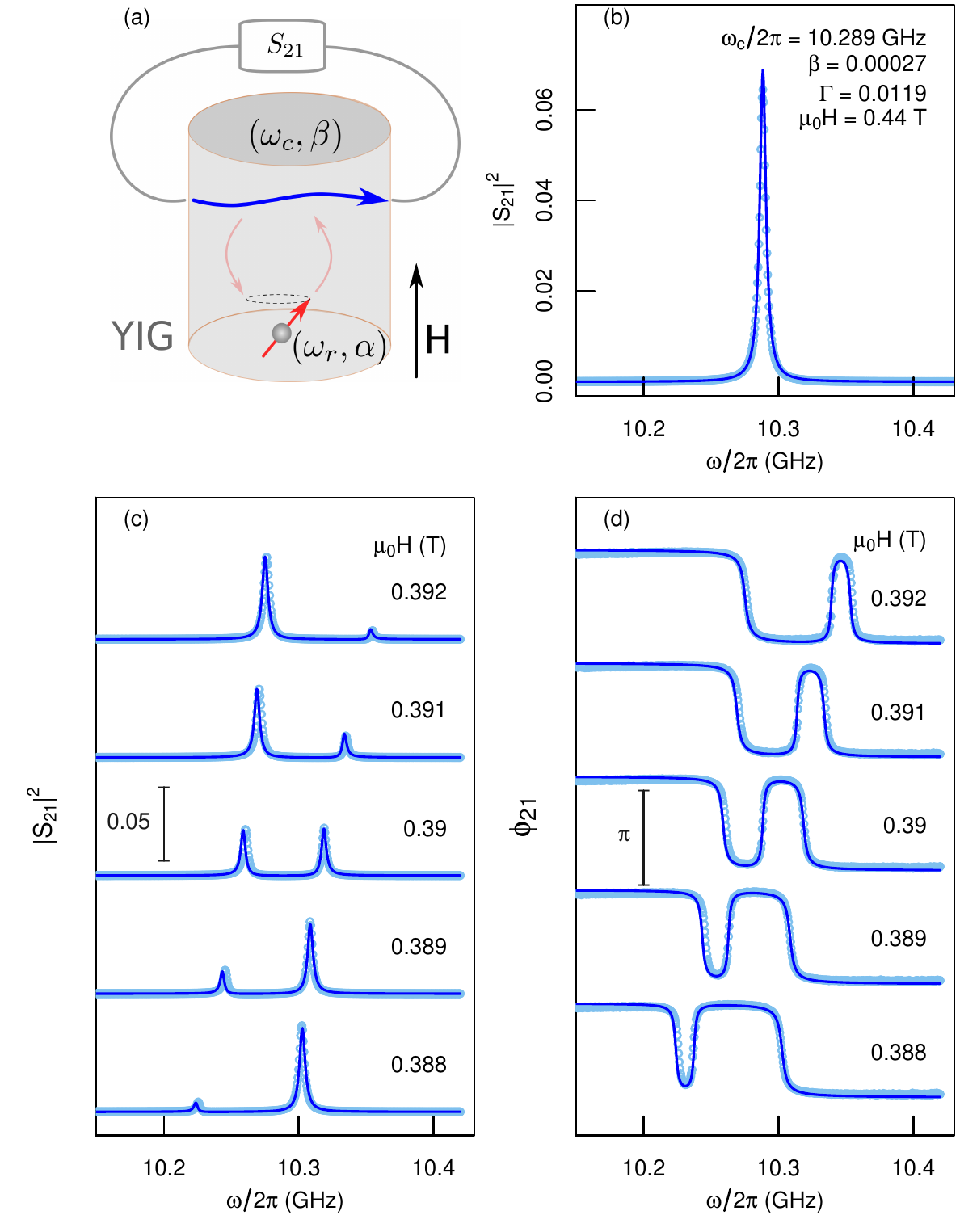}
\caption{(a) A sketch of the apparatus used to measure the microwave transmission, $S_{21}$, of a FMR-cavity mode coupled system using a VNA. (b) The cavity mode is characterized when it is uncoupled from the FMR. (c) and (d) display the amplitude and phase of the microwave transmission $S_{21}$ at different external magnetic fields in the coupled system.}
\label{fig1}
\end{figure} 

In our experiment we used a cylindrical microwave cavity made of oxygen-free copper with a diameter and height of 25 and 32.2 mm respectively, as sketched in Fig. \ref{fig1} (a).
Two coaxial cables connected from the VNA to the cavity enabled microwave transmission measurements, with an impedance factor $\Gamma$ characterizing the cavity/cable impedance mismatch.
A microwave signal $h_{0}$ with frequency $\omega$ drives the cavity, and the response function of the empty cavity can be written as \cite{Harder2016} $S_{21} =  \Gamma \frac{\omega^{2}}{\omega^{2} - \omega_{c}^{2} + 2i\beta\omega_{c}\omega} = \Gamma h/h_{0}$.  Here $\omega_{c}$ is the geometry dependent cavity resonance frequency, $\beta$ is the cavity damping and $h$ is the microwave magnetic field inside of the cavity which is supplied by the driving field $h_{0}(\omega)$. 
The transmission amplitude, $|S_{21}|^{2}$, of the empty cavity is plotted in Fig. \ref{fig1} (b) and shows a sharp peak at the cavity resonance. 
From the empty cavity measurement we determined the cavity mode frequency $\omega_{r}/2\pi$ = 10.289 GHz, damping $\beta$ = 0.00027 and impedance matching $\Gamma$ = 0.0119.

To observe spin-photon coupling a 1 mm diameter YIG sphere  \footnote{\url{http://www.ferrisphere.com}} was placed inside of the cavity.
Since the uncoupled FMR resonance frequency follows the Kittel dispersion, $\omega_{r} = \gamma(H + H_\text{A})$, the application of an external magnetic field $H$ enables tuning of the FMR frequency, which can be brought into coincidence with the cavity mode in order to observe strong coupling.
In our sample the YIG gyromagnetic ratio and anisotropy field were found to be $\gamma$ = 28$\times 2\pi~\mu_{0}$GHz/T and $\mu_0H_\text{A} = 0.0225$ T respectively.  

The YIG FMR is driven by the local microwave magnetic field according to the Landau-Lifshitz-Gilbert (LLG) equation, $m = \frac{\omega_{m}\kappa h}{\omega - \omega_{r} + i\alpha\omega}$, which describes the magnetization dynamics in both the strong and weak coupling regimes.  Here the Gilbert damping constant $\alpha$ and the saturation magnetization $M_{0}$ ($\omega_{m} = \gamma M_{0}$) are properties of the FMR, while the dimensionless parameter $\kappa$ characterizes the driving efficiency of the local microwave field.  In the case of strong coupling, $\kappa$ measures the FMR/cavity coupling strength.  
 
The amplitude of the microwave transmission $|S_{21}|^{2}$ is plotted as the symbols in Fig. \ref{fig1} (c) as a function of microwave frequency at various external magnetic fields.  The two resonance peaks correspond to the hybridized modes of the coupled system.  A key signature of the coupling is an anti crossing of the hybridized modes which is shown in Fig. \ref{fig1} (c) and which has been widely studied by many groups \cite{Huebl2013a, Zhang2014, Bai2015}.  In addition to the amplitude of the complex microwave transmission, the phase of $S_{21}$, $\phi_{21}$, is plotted as symbols in Fig. \ref{fig1} (d).  Two $\pi$-phase-delays can be observed, corresponding to the two normal mode resonance peaks in Fig. \ref{fig1} (c).
An additional opposite $\pi$-phase-shift is observed between the two normal modes.  This additional shift corresponds to the anti-resonance, as we will discuss below.

\section{Determination of the Dynamic Magnetization Phase}

Previous work has shown that the strong spin-photon coupling can be accurately described by the coupled Maxwell and LLG equations \cite{Bai2015, Harder2016},
\begin{equation}
\begin{pmatrix}
\omega^{2} -\omega_{c}^{2} + 2i\beta\omega_{c}\omega & \kappa\omega^{2} \\
\kappa\omega_{m} & \omega-\omega_{r} +i\alpha\omega 
\end{pmatrix}\begin{pmatrix}
h\\
m
\end{pmatrix} = \begin{pmatrix}
\omega^{2}h_{0}\\
0
\end{pmatrix},
\label{Classicalmodel}
\end{equation}
where the first equation describes the cavity mode and boundary conditions of Maxwell's equations using a model RLC circuit, and the second equation describes the magnetization dynamics according to the LLG equation as discussed above.  The dimensionless parameter $\kappa$ denotes the FMR/cavity coupling strength and can be directly related to the microscopic properties of the system \cite{Harder2016}.  The coupled system is driven by the microwave magnetic field $h_{0}(\omega)$ supplied by the VNA, and the transmission of the coupled system is determined from Eq.~\eqref{Classicalmodel},
\begin{align}
S_{21} &=\Gamma h/h_{0} \nonumber \\  &= \Gamma \frac{(\omega-\omega_{r}+ i\alpha\omega)\omega^{2} }{(\omega^{2} - \omega_{c}^{2} + 2i\beta\omega_{c}\omega)(\omega - \omega_{r}+i\alpha\omega) - \kappa^{2}\omega_{m}\omega^{2}}.
\label{responsefunction01}
\end{align}
The amplitude and phase of the microwave transmission calculated according to Eq.~\eqref{responsefunction01} are plotted as solid lines in Fig. \ref{fig1} (c) and (d) respectively and agree well with the experimental results.  From these fits a coupling strength of $\kappa = 0.0059$ was determined.

An important consequence of Eq.~\eqref{Classicalmodel} is that it enables the determination of the dynamic magnetization phase from the measured microwave transmission phase.  To understand this, we first note that since the magnetic field ($h$) and magnetization ($m$) defined by Eq.~\eqref{Classicalmodel} are complex valued, we may define the phases $\phi_{h}$ and $\phi_{m}$ respectively (to set these phases we take the phase of $h_{0}$ as a reference).  Now from Eq.~\eqref{Classicalmodel} it is clear that $\phi_m$ and $\phi_h$ are not independent but are related according to,  
\begin{equation}
\phi_{m} = \phi_{h} + \text{arccot}\left( \frac{\omega -\omega_{r}}{\alpha\omega} \right).
\label{phasem01}
\end{equation}
Eq.~\eqref{phasem01} makes it clear that the phase relation between the dynamic magnetization and the microwave field is solely determined by the YIG FMR in terms of $\omega_{r}$ and $\alpha$.  This means that Eq.~\eqref{phasem01} enables the determination of $\phi_m$ from $\phi_h$ once $\omega_{r}$ and $\alpha$ have been properly characterized.  Finally, since the microwave transmission is proportional to the microwave magnetic field $h$ regardless of whether the system is strongly coupled, the phase $\phi_{21}$ for both strongly and weakly coupled systems should directly represent the phase of the microwave magnetic field $\phi_{h}$ and therefore $\phi_{21} \equiv \phi_h$.  For this reason we have relabelled the axis between Fig. \ref{fig2} (d) and Fig. \ref{fig3} (d) for added clarity.  This means that we can determine the dynamic magnetization phase $\phi_m$ from the experimentally measured microwave transmission phase $\phi_{21}$.  This understanding is important in order to properly analyze the coherence of coupled spin-photon systems, for example to understand the phase information of magnon dark mode systems, and is related to anti-resonance phenomena in that a phase shift in both $\phi_m$ and $\phi_h$ is observed at anti-resonance, as discussed next.
 
\section{Anti-Resonance and Phase Correlation Due to Coupling}

\begin{figure}[tb]
 \includegraphics[width = 7.70 cm]{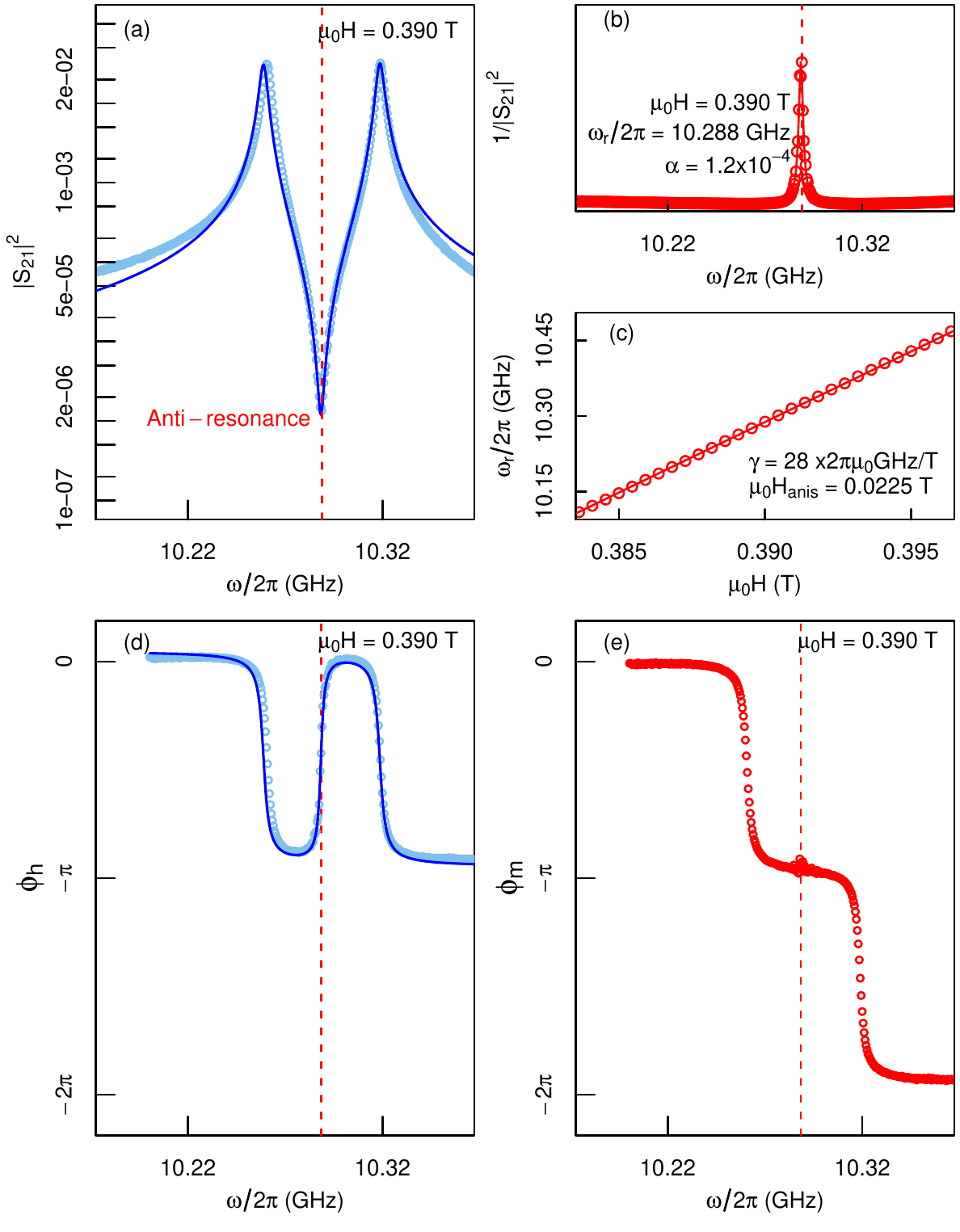}
\caption{(a) An anti-resonance, highlighted by the red dashed line, was observed in the amplitude of the microwave transmission $S_{21}$ at $\mu_{0}H$ = 0.390 T. (b) The inverse of the anti-resonance signal can be fit using Eq.~\eqref{anti-resonance} to find the uncoupled FMR frequency and damping coefficient, which enables (c) a determination of the uncoupled FMR $\omega_{r} -H$ dispersion. The solid curves in (a) and (c) are calculation results from Eq.~\eqref{responsefunction01}. (d) The measured microwave transmission phase which can be used to calculate (e) the dynamic magnetization phase $\phi_{m}$ using Eq.~\eqref{phasem01}.}
\label{fig2}
\end{figure} 

Fig. \ref{fig2} (a) displays the microwave transmission amplitude $|S_{21}|^{2}$ on a logarithmic scale at a fixed magnetic field $\mu_{0}H$ =  0.390 T using the same data as shown in Fig. \ref{fig1} (c).  On this scale the amplitude minimum at $\omega = \omega_{r}$, highlighted by the vertical dashed line, is readily observed.  This anti-resonance feature corresponds to the minimum of Eq.~\eqref{responsefunction01}, which can be explicitly seen by examining the inverse of the transmission near $\omega_{r}$, 
\begin{equation}
\frac{1}{S_{21}}  \propto \frac{1}{\omega -\omega_{r} + i\alpha\omega}.
\label{anti-resonance}
\end{equation}
Interestingly, Eq.~\eqref{anti-resonance} indicates that the anti-resonance frequency only depends on the uncoupled FMR frequency $\omega_{r}$, which differs from the FMR/cavity normal mode frequencies which depend strongly on the coupling strength. 
The inverse of the microwave transmission in Fig. \ref{fig2} (a) is plotted in Fig. \ref{fig2} (b) and was fit using Eq.~\eqref{anti-resonance}.
From this fit the uncoupled FMR was found to have a resonance frequency $\omega_{r}/2\pi$ = 10.288 GHz and a Gilbert damping $\alpha$ = 0.00012. 
The uncoupled FMR resonance position is plotted as a function of the external magnetic field in Fig. \ref{fig2} (c) and fit according to $\omega_{r} = \gamma(H + H_\text{A})$ to determine the gyromagnetic ratio $\gamma = 28\times 2\pi ~\mu_{0}$GHz/T and the anisotropy field $\mu_{0}H_\text{A}$ = 0.00225 T.  Therefore examination of the anti-resonance enables the independent characterization of the uncoupled FMR subsystem, even though the FMR/cavity system is strongly coupled.

Fig. \ref{fig2} (d) shows the measured microwave field phase with the vertical dashed line highlighting another important feature of the anti-resonance -- this position corresponds to an additional $\pi$-phase-shift, opposite to the phase shift induced at resonance. 
Using the uncoupled FMR frequency and the Gilbert damping determined using the anti-resonance, the dynamic magnetization phase $\phi_{m}$ can be obtained according to Eq.~\eqref{phasem01} as shown in Fig. \ref{fig2} (e).  Comparing Fig. \ref{fig2} (d) and (e), the dynamical magnetization phase $\phi_{m}$ is in phase with the microwave magnetic field phase $\phi_{h}$ at frequencies below $\omega_{r}$ and out-of-phase by $\pi$ at frequencies higher than $\omega_{r}$.  $\phi_{m}$ is delayed twice by the two normal modes produced by the coupling.  Meanwhile, the phase of the microwave transmission has an additional opposite phase jump due to the anti-resonance at $\omega_{r}$.  Therefore insight from the anti-resonance features both enables the determination of the phase evolution by determining $\omega_r$ and $\alpha$, and also clarifies the origin of the ``phase jump" frequencies.
 
\section{Correlated Phases of a Magnon Dark Mode} 
  
As an application of the anti-resonance analysis procedure and the phase characterization we have presented above, we now characterize the phase information of a magnon dark mode system originally proposed by Zhang et al. for use as a gradient memory architecture \cite{Zhang2015g}.  This system consists of two YIG spheres strongly coupled to a single cavity mode.  By tuning the relative phase between magnetization and microwave magnetic field via a static field gradient applied between the two YIG spheres, a magnon dark mode can be switched between bright and dark states.  However in order to actually function as a memory, the phase information of the dark mode system should be characterized, which until now was not possible.
 
The two YIG/one cavity mode system can be understood based on a simple extension of the classical model described by Eq.~\eqref{Classicalmodel}.  The addition of another YIG mode coupled to the cavity yields, 
\begin{widetext}
\begin{equation}
\begin{pmatrix}
\omega^{2} -\omega_{c}^{2} + 2i\beta\omega_{c}\omega & \kappa_{1}\omega^{2} & \kappa_{2}\omega^{2} \\
\kappa_{1}\omega_{m} & \omega-\omega_{r1} +i\alpha\omega  &0 \\
\kappa_{2}\omega_{m} & 0 & \omega-\omega_{r2} +i\alpha\omega
\end{pmatrix}\begin{pmatrix}
h\\
m_{1}\\
m_{2}
\end{pmatrix} = \begin{pmatrix}
\omega^{2}h_{0}\\
0\\
0
\end{pmatrix}
\label{Classicalmodel03}
\end{equation}
\end{widetext}
where, $\omega_{r1, r2}$ are the two uncoupled FMR frequencies, $\kappa_{1,2}$ are the two FMR/cavity coupling strengths and we define the microwave cavity magnetic field phase and the dynamical magnetization phases to be $\phi_{h}$ and $\phi_{1,2}$ respectively.
Rather than applying a local magnetic field to individually tune the FMR frequency, we use two YIG spheres which have slightly different anisotropy fields $\mu_{0}H_\text{A}$ (0.00925 T and 0.00225 T respectively), which leads to the spheres having different resonance frequencies at a given external magnetic field. 
The microwave transmission due to the three mode coupled system is determined by Eq.~\eqref{Classicalmodel03},
\begin{widetext}
\begin{equation}
S_{21}  = \Gamma \frac{(\omega-\omega_{r1}+ i\alpha\omega)(\omega-\omega_{r2}+ i\alpha\omega)\omega^{2} }{(\omega^{2} - \omega_{c}^{2} + 2i\beta\omega_{c}\omega)(\omega - \omega_{r1}+i\alpha\omega)(\omega - \omega_{r2} +i\alpha\omega) - \kappa_{1}^{2}\omega_{m}\omega^{2}(\omega - \omega_{r2} +i\alpha\omega) - \kappa_{2}^{2}\omega_{m}\omega^{2}(\omega - \omega_{r1} + i\alpha\omega)}.
\label{responsefunction02}
\end{equation}
\end{widetext}
From Eq.~\eqref{responsefunction02} it is clear that there are now two anti-resonances of the microwave transmission, at $\omega = \omega_{r1}$ and $\omega = \omega_{r2}$.

\begin{figure}[!ht]
 \includegraphics[width = 7.70 cm]{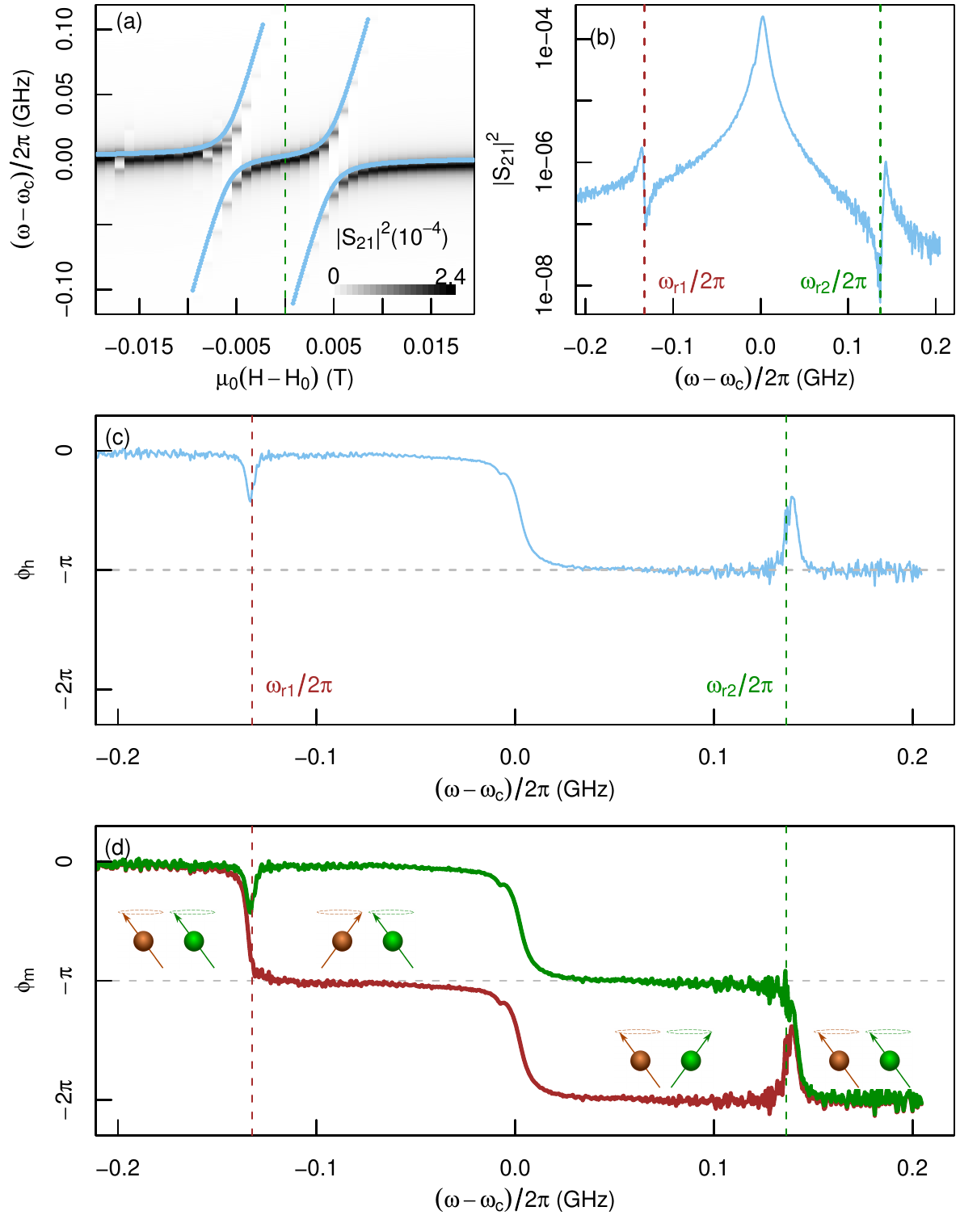}
\caption{(a) Amplitude of the microwave transmission $|S_{21}|^{2}$ as a function of the microwave frequency and the magnetic field for the microwave cavity coupling with two YIG spheres. At the external magnetic field indicated by the vertical line, the transmission amplitude $|S_{21}|^{2}$ was plotted in (b) and shows two anti-resonance frequencies. (c) The measured microwave transmission phase has two opposite phase jumps due to the two anti-resonances. (d) The phases of the dynamical magnetization for each YIG FMR are calculated. The inset sketch shows that the two FMRs are out of phase between the two anti-resonance frequencies ($\omega_{r1}$ and $\omega_{r2}$).}
\label{fig3}
\end{figure}

The microwave transmission was measured as a function of microwave frequency $\omega$ and external magnetic field $H$ and is plotted in Fig. \ref{fig3} (a), where the solid curves are roots of the determinant in Eq.~\eqref{Classicalmodel03}.  At a given external magnetic field the amplitude of the microwave transmission, shown in Fig. \ref{fig3} (b), has three normal modes, which can be seen by the 3 peaks.  The two dips indicate two anti-resonances corresponding to the two uncoupled FMR frequencies, one from each YIG sphere, and are labelled by the vertical dashed lines $\omega_{r1, r2}$.  Both YIG samples have a similar Gilbert damping constant, $\alpha = 1 \times 10^{-4}$, determined from an anti-resonance analysis.  The presence of two anti-resonances is also confirmed by two positive phase jumps observed in the microwave magnetic field phase, $\phi_{h}$, which can be seen in Fig. \ref{fig3} (c) in addition to the three $\pi$-phase delays induced by the normal mode resonance. 

The two magnetization phases are related to the microwave magnetic field phase, in analogy with Eq.~\eqref{phasem01}
\begin{subequations}
\begin{align}
\phi_{m1} &= \phi_{h} + \text{arccot}\left( \frac{\omega -\omega_{r1}}{\alpha\omega} \right) \\
\phi_{m2} &= \phi_{h} + \text{arccot}\left( \frac{\omega -\omega_{r2}}{\alpha\omega} \right).
\end{align}
\label{phasem02}%
\end{subequations}
Following the procedure we have described above, each magnetization phase was evaluated and are plotted in Fig. \ref{fig3} (d), where the relative phase between the two FMRs are indicated by the insets.  Before the first anti-resonance frequency $\omega_{r1}$, the magnetization of both FMRs are in phase with the microwave magnetic field.
In between $\omega_{r1}$ and $\omega_{r2}$ the magnetizations of the two 
YIG spheres are out-of-phase by $\pi$ with each other, forming a magnon dark mode (which is visible here since the anisotropy fields are chosen to be distinct).   However even within this range, while remaining out of phase with one another, the YIG magnetizations both experience a $\pi$ phase shift due to a normal mode near the cavity frequency $\omega_{c}$.  Finally, after the second anti-resonance frequency $\omega_{r2}$, the magnetization from both FMRs are still in phase with each other but out of phase by $\pi$ with the microwave magnetic field.  Therefore a combination of anti-resonance and phase analysis enables a characterization of the magnon dark mode system, and in particular directly reveals the in/out of phase properties which enable the formation of dark modes. 

\section{Conclusions}

We measured a strongly coupled FMR/cavity system using microwave transmission, observing an anti-resonance which is solely determined by the FMR in YIG rather than the normal modes produced by coupling.  Therefore an analysis of the anti-resonance enables characterization of the FMR subsystem, despite the fact that it is strongly coupled to the cavity mode.  Such an analysis enables the determination of the FMR frequency $\omega_r$ and Gilbert damping $\alpha$, which can in turn be used to analyze the magnetization phase.  In this way we have demonstrated a method to analyze the phase of each subsystem in a strongly coupled FMR/cavity system, and the resulting phase information can be understood in terms of the presence of resonance and anti-resonance frequencies.  This method has been applied to a magnon dark mode system, enabling a direct determination of the phase evolution which results in the formation of a dark mode.  Our work therefore provides a standard procedure for the analysis of the individual subsystems and their phase evolution in a strongly coupled system. 
\vspace{20pt}
\section*{Acknowlegements}
M. H. is supported by an NSERC CGSD Scholarship.  P. H. is supported by the UMGF program. This work was funded by NSERC, CFI, and NSFC (No. 11429401) grants (C.-M. H.). 
\vspace{120pt}
\bibliography{mainText.bbl}
%\bibliography{/users/michaelharder/desktop/bibliography/library.bib}

\end{document}